\documentclass[aps,prl,twocolumn,showpacs,preprintnumbers,superscriptaddress,amsmath,amssymb,prl]{revtex4-1}

\usepackage{graphicx}
\usepackage{dcolumn}
\usepackage{bm}
\usepackage{hyperref}
\usepackage{color}
\usepackage{natbib}
\usepackage[OT1]{fontenc}

\begin{document}

\title{Correlated Insulator Collapse due to Quantum Avalanche via
In-Gap Ladder States}

\author{Jong E. Han}
\email{jonghan@buffalo.edu}
\affiliation{Department of Physics, State University of New York at Buffalo, Buffalo, New York 14260, USA}
\author{Camille Aron}
\affiliation{Laboratoire de Physique de l'\'Ecole Normale Sup\'erieure, ENS, Universit\'e PSL, CNRS, Sorbonne Universit\'e, Universit\'e Paris Cit\'e, F-75005 Paris,  France}
\affiliation{Institute of Physics, \'Ecole Polytechnique F\'ed\'erale de Lausanne (EPFL), CH-1015 Lausanne, Switzerland.}
\author{Xi Chen}
\affiliation{Department of Physics, State University of New York at Buffalo, Buffalo, New York 14260, USA}
\author{Ishiaka Mansaray}
\affiliation{Department of Physics, State University of New York at Buffalo, Buffalo, New York 14260, USA}
\author{Jae-Ho Han}
\affiliation{Center for Theoretical Physics of Complex Systems, Institute for Basic Science(IBS), Daejeon 34126, South Korea}
\author{Ki-Seok Kim}
\affiliation{Department of Physics, POSTECH, Pohang, Gyeongbuk 37673, South Korea}
\author{Michael Randle}
\affiliation{Department of Electrical Engineering, State University of New York at Buffalo, Buffalo, New York 14260, USA}
\author{Jonathan P. Bird}
\affiliation{Department of Physics, State University of New York at Buffalo, Buffalo, New York 14260, USA}
\affiliation{Department of Electrical Engineering, State University of New York at Buffalo, Buffalo, New York 14260, USA}

\date{\today}

\begin{abstract} 

The significant discrepancy observed between the predicted and
experimental switching fields in correlated insulators under a DC
electric field far-from-equilibrium necessitates a reevaluation of
current microscopic understanding.  Here we show 
that an electron avalanche can occur in the bulk limit of such insulators at arbitrarily
small electric field by introducing a generic model of electrons coupled
to an inelastic medium of phonons.
The quantum avalanche arises by the generation of
a ladder of in-gap states, created by a multi-phonon emission process.
Hot-phonons in the avalanche trigger a premature and partial collapse of
the correlated gap. The phonon spectrum dictates the
existence of two-stage versus single-stage switching events which we
associate with charge-density-wave and Mott resistive phase transitions, respectively.
The behavior of electron and phonon temperatures, as
well as the temperature dependence of the threshold fields, demonstrates
how a crossover between the thermal and quantum switching scenarios
emerges within a unified framework of the quantum avalanche.

\end{abstract}

\maketitle

In spite of recent progress in our understanding of the nonequilibrium
many-body state of matter, one of the long-standing problems that has
remained unresolved concerns the microscopic mechanism behind the
insulator-to-metal transition (IMT) of strongly-correlated electronic
systems, driven by a DC electric field. For more than five decades, the
community has fiercely debated the origin of the dielectric breakdown in
charge-density-wave (CDW)
systems~\cite{bardeenPT,bardeen1989,ong1979,grunerRMP,maki1977,thornehistory},
for which the reported threshold electric fields are orders of magnitude
smaller than theoretical estimates based on the
Landau-Zener~\cite{zener} mechanism. In the late seventies, this problem
led to the formulation of the classical theory of depinning, in which
the CDW order parameter is understood as being pinned by the presence of
disorder and can be abruptly unpinned under the action of a static
electric field~\cite{fukuyama1978,lee1979,fisherCDW}. More recently, a
similar mismatch between theory and experiments has also been found in
studies of transition-metal compounds such as Mott
insulators~\cite{janod,stoliar,guiot,Ridley,sblee,zhang_nano}. In spite
of the emerging potential of these materials for applications such as
non-volatile neuromorphic computing~\cite{del_valle}, the lack of
understanding of the microscopic origins of their resistive transitions
is a bottleneck to the development of such technologies.

From the early days of CDW research, the theoretical paradigm for the
resistive transition in correlated systems has been the
Landau-Zener mechanism~\cite{ong1979,zener}. This model predicts switching
fields on the order of
\begin{equation}
E_{\rm LZ}\sim \frac{\Delta^2}{e\hbar v_F},
\label{eimt0}
\end{equation}
with the (zero-field) insulating gap $\Delta$ and the Fermi-velocity
$v_F$. Using typical electronic energy scales in the
above expression, we obtain a rough estimate of $E_{\rm LZ}\sim 10^{-2}$
V/\AA\ $=10^6$ V/cm, many orders of magnitude larger than the switching
fields of $\lesssim$ 10 kV/cm found in Mott insulators~\cite{janod} and
$1-10^3$ V/cm in CDW
materials~\cite{bardeenPT,bardeen1989,ong1979,grunerRMP}. Over the
years, various theoretical attempts have been made to improve the
description of the resistive switching transition in a many-body
context. Existing theories include the explicit time-evolution of the
Hubbard chain~\cite{oka2003}, the multi-band Hubbard model~\cite{mazza},
disorder-driven ~\cite{sugimoto} and nonequilibrium phase
transitions~\cite{han2018}, spatial inhomogeneity~\cite{nanolett}, and
Coulomb blockade of multi-solitons~\cite{miller2000,maki1977}.
While bandedge
broadening~\cite{franzkeldysh,davies,han2018} and metal-insulator
filament dynamics~\cite{nanolett} by a uniform electric field are known
to reduce the switching fields, these effects are quite modest.
Recently, a nucleation mechanism for metallic domains, assisted by
electron-phonon coupling, has been reported~\cite{zhang_chern}.  These
various microscopic models have been unable to settle the aforementioned
energy-scale problem, or to provide clarity to the age-old debate over
the role of thermal~\cite{zimmers} versus
electronic~\cite{giorgianni2019,jager} mechanisms in the resistive
transition. It has therefore long been speculated that an important
ingredient must be missing.
 
With a wide class of switching materials~\cite{janod}
the mechanisms could be just as diverse. The impact ionization in
semiconductors and the electro-migration in nanoscale oxide devices have
been well studied. In correlated oxides much of the phenomenology has
been understood in terms of the filamentary
dynamics~\cite{stoliar,janod,Ridley,zimmers}. The recent report
on the nano-scale resolution of non-filamentary patterns in
Ca$_2$RuO$_4$~\cite{zhang_nano}, however, suggests quite a different
non-thermal mechanism, reminiscent of the switching in organic
solids~\cite{kumai_science} in which perpendicular patterns to bias have
been observed.  The importance of the pattern formation along the bias
direction in device geometry has been pointed out by recent
theories~\cite{ribeiro,tanaka_crossover,dutta_spatial}. Given the
diverse switching phenomena, we limit the scope of this paper to a new
switching mechanism in the bulk correlated insulators where the
switching is understood as phase transition controlled by the electric
field.

The colossal mismatch of the switching fields in the field-driven Mott
and CDW transitions motivates us to look for a common underlying
mechanism that is shared between them.  Despite the differences in CDW
and Mott phenomenology, it is often hard to disentangle one mechanism
from the other~\cite{zhuCDW}.  Here we propose
 that the resistive transitions
in CDW and Mott insulators can be coherently explained in terms of the role of
inelastic many-body processes. We show how these phonon-emitting
transitions can lead to the generation of in-gap states, resulting in a
quantum avalanche and a destruction of the correlated gap, at much
smaller electric field scales than previously predicted. 

\begin{figure}
\begin{center}
\rotatebox{0}{\resizebox{3.5in}{!}{\includegraphics{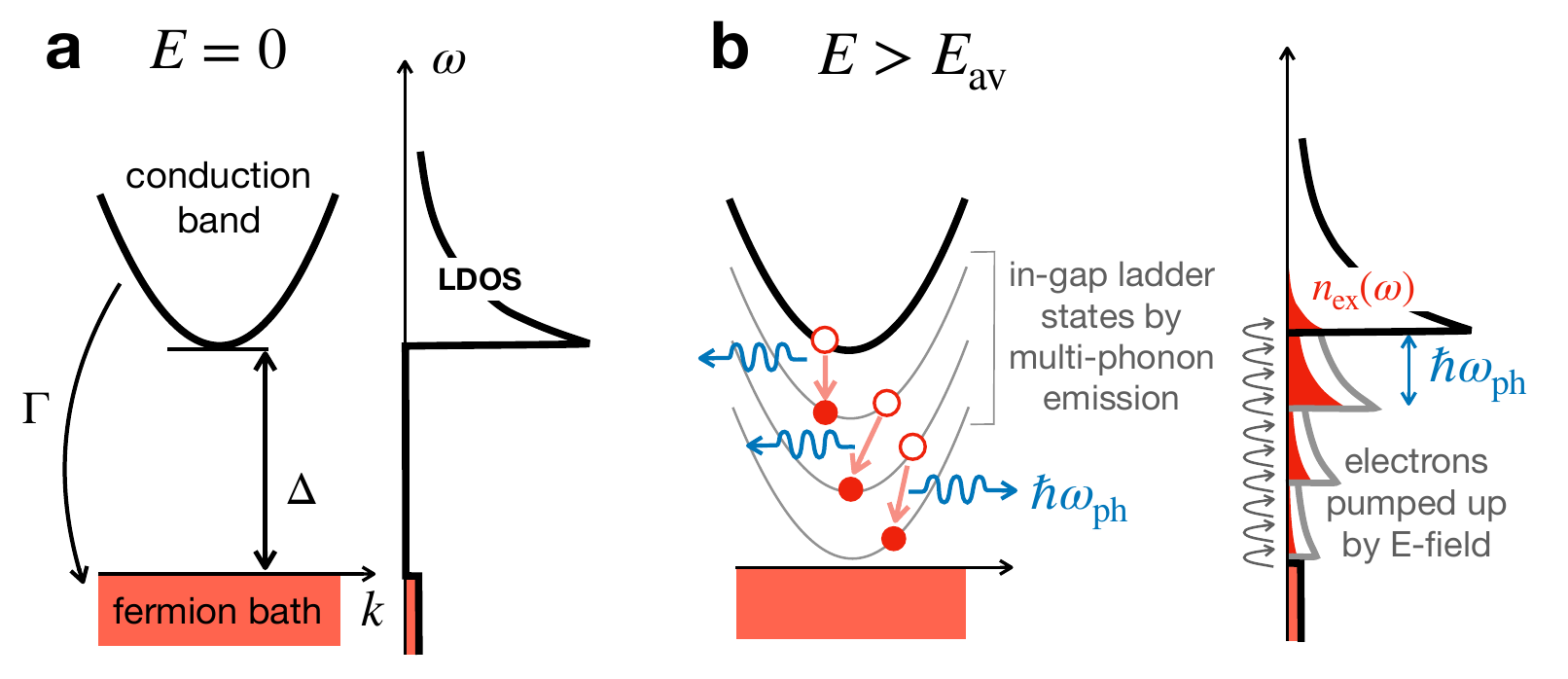}}}
\caption{{\bf Avalanche Mechanism a,} 
Zero electric field, $E = 0$: (left) Spectral gap $\Delta$ between the electronic band and the Fermi level.
The hybridization to a fermionic bath
provides a finite decay rate $\Gamma$.
(right) The local density of
states (LDOS) for electrons that occupy only bath states (red shading).
{\bf b,} (left) Electronic ladder-like states after successive phonon-emission events
(blue) form quantum pathways for avalanche at $E>E_{\rm
av}$. (right)
Once the in-gap ladder states are formed, electrons are
pumped through them into the conduction band by the
electric-field. A finite density of electronic excitation
$n_{\rm ex}(\omega)$ at energy $\omega$ (red shading) now populates the in-gap LDOS.
}
\label{fig1}
\end{center}
\end{figure}

For a conceptual understanding of the avalanche mechanism, we start by
considering a one-dimensional conduction band of mass $m$ separated by a
gap $\Delta$ from the Fermi energy, see Fig.~\ref{fig1}~{\bf a}.  The
electrons are subject to three processes.  First, they are accelerated
by a DC electric field $E$.  Second, they can emit optical phonons of
energy $\hbar \omega_{\rm ph}$ by means of an electron-phonon scattering
process controlled by the coupling parameter $g_{\rm ep}$. Finally,
dephasing introduces a finite lifetime $\hbar/\Gamma$ to the electrons,
setting a cutoff on the time scale over which they can form the
multi-phonon state.  The dephasing may arise from electron tunneling to
a substrate, virtual transitions to higher electronic states, and other
many-body mechanisms separate from the electron-phonon interaction.
Hereafter, we adopt a unit system in which $\hbar=1$, the electric
charge $e=1$, the Boltzmann constant $k_B=1$, and the lattice constant
$a=1$.

Let us now give a back-of-the-envelope criterion for an avalanche that
can be triggered by a combination of the three aforementioned electronic
processes. We note that, in an insulator subject to an external bias,
there will exist dilute, yet nonetheless finite, charge excitations.
Due to the Franz-Keldysh effect~\cite{franzkeldysh}, the
bandedge smears into the gap and the transition of energetic
electrons into those states by phonon emission is substantially enhanced over that
in the equilibrium limit. Importantly, electron-phonon scattering
processes create a ladder of intermediate replica bands that are equally
spaced by $\omega_{\rm ph}$.  The minimal number of successive
phonon-scattering events to bridge the gap is $N_{\rm ladder} \sim
\Delta / \omega_{\rm ph}$, see Fig.~\ref{fig1}{\bf b}.  The timescale it
takes to accelerate an electron in a dispersive band, to gain the energy
of a single phonon, can be estimated as $\tau \sim (m \, \omega_{\rm
ph})^{1/2}/E$ with the electron band mass $m$. The electron-phonon scattering rate is roughly
proportional to the dimensionless electron-phonon mass-renormalization
factor $\lambda\equiv (g_{\rm ep}/\omega_{\rm ph})^2 \nu_0$, where the
typical density of states $\nu_0 \sim m$. The number of phonons that are
emitted during the electronic lifetime $1/\Gamma$ is then $N_{\rm ep}
\sim {\lambda}/{\Gamma \tau} $.  Consequently, the criterion for the
avalanche becomes $N_{\rm ep} = N_{\rm ladder}$, yielding the following
electric field required to trigger the avalanche
\begin{equation} \label{eq:E_av}
E_{\rm av}\sim\frac{\hbar\Gamma\Delta}{eg_{\rm
ep}^2}\left[\frac{(\hbar\omega_{\rm ph})^3}{2m}\right]^{1/2}.
\end{equation}

This outcome is highly non-trivial in that the dissipative electron
decay rate $\Gamma$ plays a crucial role in the onset of this
nonequilibrium quantum phase transition. The $\Gamma\to
0$ limit is singular in nonequilibrium. By taking $\Gamma\to 0$ before or
after the $E\to 0$ limit, we arrive at an effectively
infinite~\cite{eckstein2011} or a zero temperature
(equilibrium) limit, respectively.
This singular nature of the $\Gamma=0$ limit plays a fundamental role in
the avalanche, as will be confirmed below. The switching field is very different in nature from
that of Eq.~(\ref{eimt0}), and we note that the avalanche via inelastic
scattering is also distinct from the conventional mechanism that has
been discussed for high-field transport in
semiconductors~\cite{wolff1954}. Most importantly, as we shall see
below, the avalanche fields predicted by Eq.~(\ref{eq:E_av}) are much
smaller than those expected from Eq.~(\ref{eimt0}).

\bigskip
\textbf{\large Results}

\bigskip
\textbf{Quantum avalanche in a rigid-band model} \\
To demonstrate the existence of the proposed avalanche mechanism, we
investigate a model of a one-band tight-binding chain under a DC
electric-field $E$, in the Coulomb gauge~\cite{liprl2015,ligraphene}
\begin{equation}
H^{\rm 1D}_{0,\rm
el}=\sum_i\left[-t(d^\dagger_{i+1}d_i+d^\dagger_id_{i+1})+(2t+\Delta_0-Ex_i)d^\dagger_i
d_i\right],
\end{equation}
where $d^\dagger_i/d_i$ is the creation/annihilation operator for an
electron at site $i$, for which the site position $x_i=ia$.  $\Delta_0$
is the bare gap and $t$ is the tight-binding parameter. The electrons
are locally coupled to phonons, which are modeled by a collection of
harmonic oscillators given by the Hamiltonian
\begin{equation}
H_{0,\rm ph}=\frac12\sum_k(p_k^2+\omega_k^2\varphi_k^2),
\end{equation}
with $\varphi_k$ the amplitude, $p_k$ the momentum, $k$ the continuum
index, and $\omega_k$ the frequency of the phonon.  In this section we
consider the case of optical phonons, with $\omega_k=\omega_{\rm ph}$,
deferring a discussion of acoustic phonons until later.  The on-site
electron-phonon coupling is given by
\begin{equation}
H_{\rm ep}=g_{\rm ep}\sum_i\varphi_id^\dagger_i d_i,
\end{equation}
with the coupling constant $g_{\rm ep}$.

We use the Schwinger-Keldysh formulation of the dynamical mean-field
theory (DMFT~\cite{georges_DMFT1996,aoki,liprl2015}) which bypasses the
transient dynamics and directly yields the homogeneous nonequilibrium
steady-state of the many-body dynamics~\cite{aron2012prl,prb2013b}. The
fermion baths enter the computation of the electronic Green's function
via local retarded and lesser self-energies at site $i$ as
\begin{equation}
\Sigma^R_{0,i}(\omega)=-i\Gamma,\quad\Sigma^<_{0,i}(\omega)=2i\Gamma
f_0(\omega+Ex_i),
\end{equation}
while the Ohmic baths~\cite{weiss} enter the phonon
Green's function via local self-energies~\cite{ligraphene,moire2023} as
\begin{equation}
\Pi^R_{0,i}(\omega)=-2i\tau_P^{-1}\omega,\quad\Pi^<_{0,i}(\omega)=-4i\tau_P^{-1}\omega
n_0(\omega).
\end{equation}
In the above expressions, $f_0(\omega)=(e^{\omega/T}+1)^{-1}$ and
$n_0(\omega)=(e^{\omega/T}-1)^{-1}$ are the Fermi-Dirac and
Bose-Einstein distributions at the bath temperature $T$, respectively,
and $\tau_P$~\cite{khurgin2007} is the phonon decay time. We compute the
second-order self-energy to electrons and phonons on the same footing.
We refer the reader to the Methods Section and to the Supplementary
Information for further details on the electron-phonon calculations.

\begin{figure}
\begin{center}
\rotatebox{0}{\resizebox{3.5in}{!}{\includegraphics{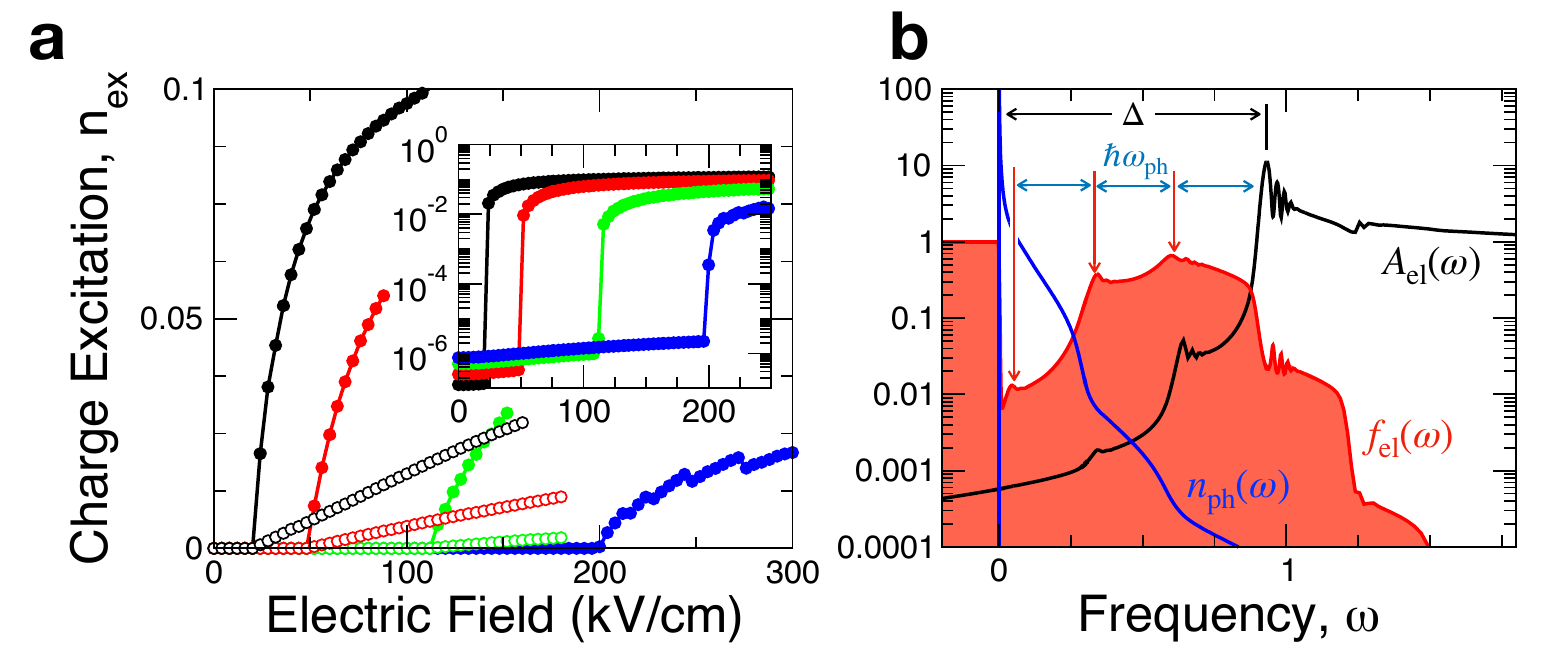}}}
\caption{{\bf Avalanche controlled by electron life-time a,}
Electron excitation per site $n_{\rm ex}$ versus
electric field $E$ at several values
of the fermion decay rate $\Gamma=1,2,4,6\times 10^{-3}$ (from left to
right). The inset shows the same data on a semi-logarithmic scale.  The
avalanche field $E_{\rm av}$ is proportional to $\Gamma$. The curves
with filled (open) symbols are computed without (with) the
nonequilibrium self-energy of the phonons.  {\bf b,} Post-avalanche
spectra of electrons and phonons, plotted using semi-log
scales. The electron spectral function $A_{\rm el}(\omega)$
with its band edge at the gap $\Delta\approx 0.9$ (slightly
reduced from $\Delta_0=1$) extends down to the reservoir level
$\omega=0$, and the distribution function $f_{\rm el}(\omega)$ reveals
in-gap state occupation with multi-phonon edges (marked by red arrows)
separated by the phonon energy $\omega_{\rm ph}$.  The distribution function $n_{\rm
ph}(\omega)$ for hot-phonons is equilibrium-like.
}
\label{fig2}
\end{center}
\end{figure}

In the calculations that follow, we choose the model parameters
$\Delta_0=1$, $t=1$, $g_{\rm ep}=0.25$, $\omega_{\rm ph}=0.3$,
$\tau_P^{-1}=0.001$, and $T=0.001$ in units of eV. We use the lattice
constant $a=5$ \AA\ to compute the electric-field. We
verify the schematic estimate of the avalanche field, Eq.~(2), via systematic comparison with
full many-body calculations over a wide range of parameters, as detailed
in the Supplementary Information. There, we demonstrate the conceptual validity of
the avalanche as arising from a competition between inelastic transport and 
dephasing and justify the limitations of Eq.~(2) in the high-field
limit. 

Fig.~\ref{fig2}{\bf a} shows the electronic
excitations $n_{\rm ex}$ per site ($0\leq n_{\rm ex}\leq 1$) when
ramping up the electric field starting from the insulator state. Filled
(empty) symbols denote results obtained by discarding (including) the
nonequilibrium self-energy of the phonons.  At the onset of the
avalanche at the threshold field $E_{\rm av}$, $n_{\rm ex}$ increases
abruptly, yet continuously, in a similar fashion as in critical
phenomena.  Notably, $E_{\rm av}$ is found to be roughly proportional to
$\Gamma$, consistent with the back-of-the-envelope estimate made in
Eq.~(\ref{eq:E_av}).  Furthermore, the pre-avalanche excitations are
inversely proportional to $E_{\rm av}$ [see inset to (a)], which
demonstrates that the avalanche is not initiated by thermal excitation
but is of quantum origin. Remarkably, the onset of the avalanche is not
affected by hot-phonon effects since they are
only infinitesimally excited at the continuous avalanche
transition, while the incoherence of phonons impedes the
avalanche resulting in a linear increase of $n_{\rm ex}$
well after the avalanche.

The nontrivial behavior of the switching field on
$\Gamma$ (see Fig.~\ref{fig2}{\bf a}) should be distinguished from
phonon-assisted tunneling~\cite{kleinman1965,vdovin2016}, which is
commonly manifested in resonant tunneling in heterostructures. With the
electron occupation quickly reaching beyond 0.1 after the avalanche, we
are in a metallic bulk limit through a phase transition, instead of in
the perturbative tunneling regime.  The avalanche shown here only
emerges after a fully self-consistent solution is reached after hundreds
of iterations, unlike what is expected from a low-order perturbative
theory~\cite{kleinman1965}. 

Figure~\ref{fig2}{\bf b} elucidates nature of the nonequilibrium state
after the avalanche.  The energy distribution $f_{\rm el}(\omega)$
reveals a sizable nonequilibrium occupation of the in-gap states around
energies at multiples of $\omega_{\rm ph}$ away from the band edge. This
confirms the involvement of multi-phonon emission in the avalanche
mechanism.  Moreover, the strong deviation of $f_{\rm el}(\omega)$ from
a Fermi-Dirac distribution points to the non-thermal nature of the
avalanche. In contrast, the phonon distribution $n_{\rm ph}(\omega)$
appears mostly thermal.

\bigskip
\textbf{Insulator-to-metal transition induced by avalanche} \\

\begin{figure*}[t]
\centering
\rotatebox{0}{\resizebox{6.5in}{!}{\includegraphics{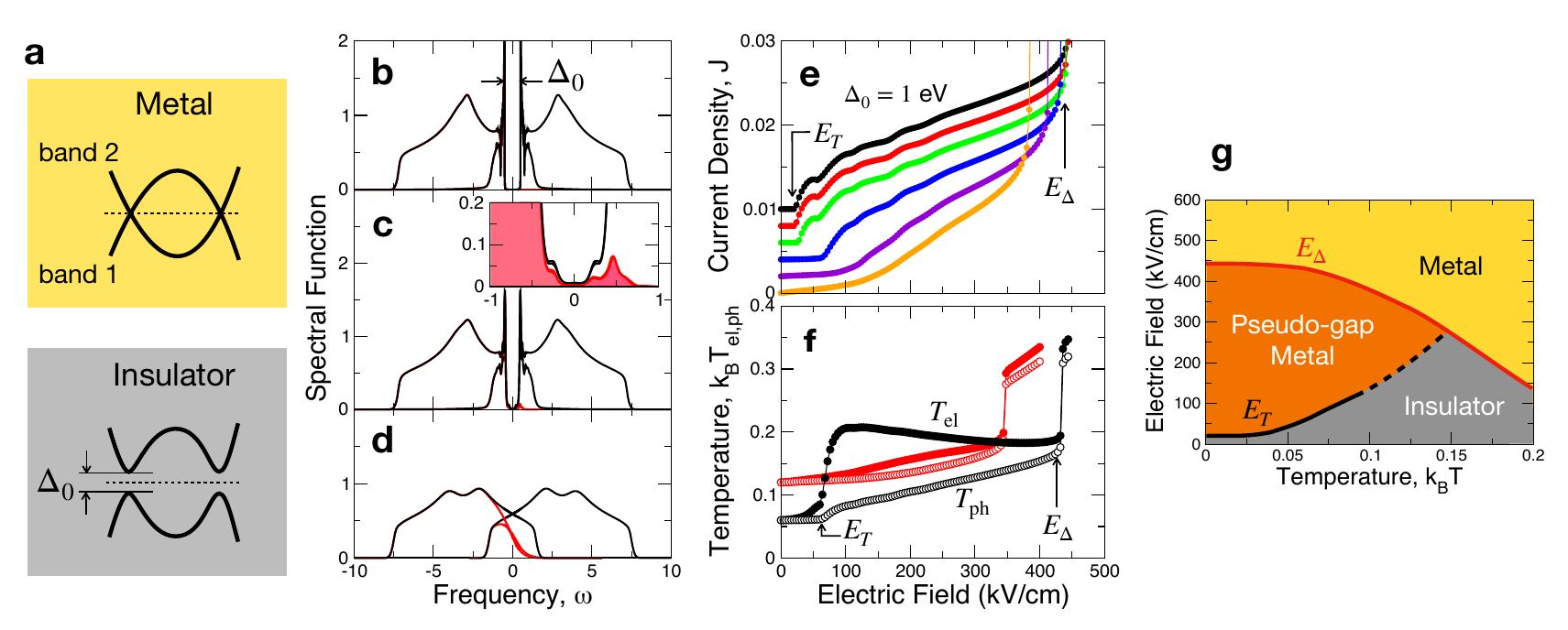}}}
\caption{
{\bf CDW-like two-stage IMT}, {\bf a,} (Top)
Band-crossing at the Fermi level (dotted line), (Bottom) Charge gap
formation due to inter-orbital interaction, {\bf b,} Density of states
at electric field $E=0$: well-defined charge gap $\Delta_0=1$.  {\bf c, } After the
first threshold $E_T$ ($E > E_{\rm T}$): pseudo-gap formation. The inset is a
close-up (solid red shading: the frequency-resolved electronic
occupation of the in-gap states).  {\bf d,} After the second threshold
$E_\Delta$ ($E > E_\Delta$): metallic density of states with a fully
collapsed gap.  {\bf e,} Current-volatge ($I-V$) characteristics for various temperatures
$T$ proceeds in two stages: a CDW-like continuous transition to a metal
at $E=E_{\rm T}$ followed by a discontinuous current jump at
$E=E_\Delta$. From top to bottom, $k_BT=0.01, 0.02,
0.04, 0.06, 0.08, 0.1$ with the curves offset in the $y$-axis by 0.002 for readability. {\bf f,} Effective temperatures $T_{\rm el,ph}$ of the
electrons (filled circle) and phonons
(empty circle), respectively. The bath temperatures are
$k_BT=0.06$ (black) and $0.12$ (red).  At $E=E_{\rm T}$, the electrons go through a
transition without heating up the phonons, while at
$E_\Delta$ the two temperatures equilibrate.  {\bf g,}
Phase diagram in the $E$--$T$ plane. The $E_{\rm T}$ and
$E_\Delta$ curves delimit a pseudo-gap metal region.
Dashed line denotes the high temperature limit where the
phase boundary is smeared. (See the main text for the parameters.)}
\label{fig3}
\end{figure*}

Having established the existence of the avalanche, we now turn our
attention to the implications of the avalanche in the context of the
resistive transition in correlated insulators.  The strong charge
fluctuations initiated by the avalanche are expected to generate phonon
excitations, which then profoundly perturb the inter-orbital mixing and
destabilize the charge gap. To address this point, we extend our model
to a two-band correlated insulator where the gap $\Delta$ between the
bands is generated by electronic interactions. (See Fig.~\ref{fig3}{\bf
a}.) Specifically, we consider the following Hamiltonian which allows us
to address both the CDW transitions and resistive switching on an equal
footing,
\begin{eqnarray}
H^{\rm 2D}_{\rm el}
& = & -t\sum_{\langle ij\rangle}\sum_{\alpha=1,2}(-1)^\alpha
(d^\dagger_{\alpha i}d_{\alpha j}+d^\dagger_{\alpha j}d_{\alpha i})
\nonumber \\
& & +\sum_{i}\sum_\alpha[(-1)^\alpha(2t-\mu)-Ex_i]d^\dagger_{\alpha i}
d_{\alpha i}
\nonumber \\
& & +\sum_i\left[ \xi(d^\dagger_{1i} d_{2i}+d^\dagger_{2i}
d_{1i})+\frac{\xi^2}{2U}\right].
\label{uterm}
\end{eqnarray}
Here, $\alpha$ is the band index and $i$ is the site index on a square
lattice with $\langle ij\rangle$ denoting nearest neighbors.  The last
term in Eq.~(\ref{uterm}) is the decoupling of an inter-orbital
Hubbard-like interaction via an auxiliary quantum field
$\xi$. The magnitude of $\xi$ respresents the order parameter that opens
a gap, i.e. the strength of the density modulation in CDW or
the charge fluctuations in correlated insulators. The phase of $\xi$
captures the phase slip in CDW materials or the Goldstone excitations in
correlated insulators. In this work, we investigate the instability of a
uniform symmetry-broken state, and ignore the phase fluctuations of
$\xi$ for simplicity. Consideration of
$\xi$-fluctuations would only weaken the order parameter and lead to an even earlier
collapse of the insulating state, and thus further supporting our
conclusions.  The static mean-field condition becomes
\begin{equation}
\xi=U\langle d^\dagger_{1i} d_{2i}+d^\dagger_{2i} d_{1i}\rangle.
\label{gapsc}
\end{equation}
The electron-phonon coupling is given by
$
H_{\rm ep}=g_{\rm ep}\sum_{i,\alpha}\varphi_{\alpha i}d^\dagger_{\alpha i} d_{\alpha i},
$
with the independent phonons coupling to each orbital.  The
fermion/phonon baths are set up as described previously with the Fermi
energy at the band crossing.

The spectrum of the phonons has important consequences for the IMT.  Let
us first discuss the electric-field driven IMT in the presence of
optical phonons with energy $\omega_{\rm ph}$, which we will associate
with the multi-stage transitions that are commonly observed in CDW
systems. Later, we will turn to the case of acoustic phonons which we
will relate to Mott systems. With increasing electric field, the
insulating system undergoes a two-stage transition to a metal as
manifested by the spectral function in Fig.~\ref{fig3}{\bf b-d}. Here,
we adjust the electron-phonon coupling $g_{\rm ep}$ such that, given
$U=2$, the initial gap $\Delta_0$ is tuned to $1.0$ at $E=0$ and at the
lowest temperature. The strength of this coupling corresponds to a
moderate (dimensionless) mass-renormalization factor of $\lambda\approx
(g_{\rm ep}/\omega_{\rm ph})^2\nu_0=0.32$ with the phonon energy
$\omega_{\rm ph}=0.3$ eV and the 2D density of states $\nu_0\approx
(8t)^{-1}$. (The damping is set to $\Gamma=\tau_P^{-1}=0.001$
eV~\cite{diaz}, and the chemical potential to $\mu=t$.) In the low-field
limit, the spectrum features a well-defined charge gap. As $E$ increases
beyond the first threshold field $E_T$, in-gap states develop via the
avalanche mechanism. This results in a pseudo-gap phase
where the system becomes metallic  while the charge gap mostly remains
intact. The energy-resolved electron occupation $n_{\rm
ex}(\omega)=(2\pi)^{-1}{\rm Im}G^<(x_i=0,\omega)$ (the red shaded area
shown in the inset to panel {\bf c}) indicates that the electric current is
carried mainly by the states of the conduction band above the gap while
the in-gap states provide the pathway for this population inversion. The
pseudo-gap regime persists until $E=E_\Delta$, when the
gap $\Delta$ collapses to zero in a strongly discontinuous transition. 

The current-voltage ($I-V$) relation in Fig.~\ref{fig3}{\bf e} shows evidence of a
two-stage IMT. First, the system continuously becomes metallic at the
threshold field $E = E_T$. Later, at the higher threshold
$E_\Delta$, the current rises discontinuously.  The
avalanche behavior discussed in Fig.~\ref{fig2} is responsible for the
threshold behavior at $E_T$.  This behavior is similar to that in CDW
systems, in which it has been widely attributed to the depinning
transition~\cite{fukuyama1978,lee1979}.  Here, we propose an alternative
mechanism of electron avalanche via inelastic scattering that does not
require any disorder or reduced dimensionality. With coupling to optical
phonons, the avalanche is not sufficiently disruptive to the gap, and
the intermediate gapped state is instead sustained over a wide range of
electric field ($E_T<E<E_\Delta$).  We note
that $E_T$ is around two orders of magnitude smaller than the switching
field expected for zero electron-phonon coupling strength
$E_\Delta(\lambda=0,\Delta_0=1)\approx 1.6$ MV/cm.

The non-thermal nature of the avalanche transition is illustrated by the
electric-field dependence of the effective temperatures for electrons
and phonons, as shown in Fig.~\ref{fig3}{\bf f}. (See Methods for the
definition of effective temperatures.) As the electric field is
increased beyond $E=E_T$, the electrons heat up immediately while the
phonons stay cold. This clearly demonstrates that heat exchange does not
trigger the avalanche. On the other hand, the full IMT at $E_\Delta$ is
initiated once the electron and phonon temperatures equilibrate,
suggesting that this second transition can be described in terms of a
thermal mechanism. It is remarkable to demonstrate that
the mechanisms for electronic and thermal switching, the topic of
intense discussions in the literature, are not exclusive of each other
but can arise simultaneously from a single microscopic model.

The phase diagram defined by the two switching fields ($E_T$ and
$E_\Delta$) is plotted in Fig.~\ref{fig3}{\bf g}. The
most notable observation here is that $E_T$ remains constant near
zero temperature, but increases at higher temperatures until it merges
with $E_\Delta$.  This may seem counter-intuitive,
since the gap is expected to decrease with increasing $T$. The gap,
however, remains nearly constant until close to a critical temperature
$T_c$, meaning that a thermal argument is not applicable.  We find that,
as suggested by the hot-phonon effects apparent in Fig.~\ref{fig2},
thermal decoherence may be detrimental to the avalanche, meaning that a
stronger electric field is required to induce an avalanche. This is a
further evidence that the threshold behavior is of quantum-mechanical
origin. In contrast to $E_T$, $E_\Delta$ decreases with
$T$, and is conventional and thermally driven.

The existence of bias-driven multiple-stage transitions in CDW systems
has been intensely
debated~\cite{bardeenPT,zaitsev-zotov2004,zaitsev-zotov2001,itkis1990,fleming1980}.
For instance, recent ARPES studies on the NbSe$_3$
system have shown unambiguously that the various CDW gap energies are
constant (marginally increasing) with increasing temperature and
observable even beyond their respective $T_c$
values~\cite{nicholson2017,nicholson2020}. This suggests a nonthermal
mechanism of gap formation below $T_c$. At temperatures beyond $T_c$,
the gap gradually closes suggesting a thermal mechanism, where these
predictions show an interesting parallel to our results. What our model
as a conceptual framework showed are that the nature of the initial
threshold~\cite{itkis1990} in the low field limit is quantum in the
sub-gap energy scale, and that, after an intermediate pseudo-gapped
phase, there are subsequent discontinuous resistive transitions that
thermally destroy the order parameter.

\begin{figure}
\begin{center}
\rotatebox{0}{\resizebox{3.5in}{!}{\includegraphics{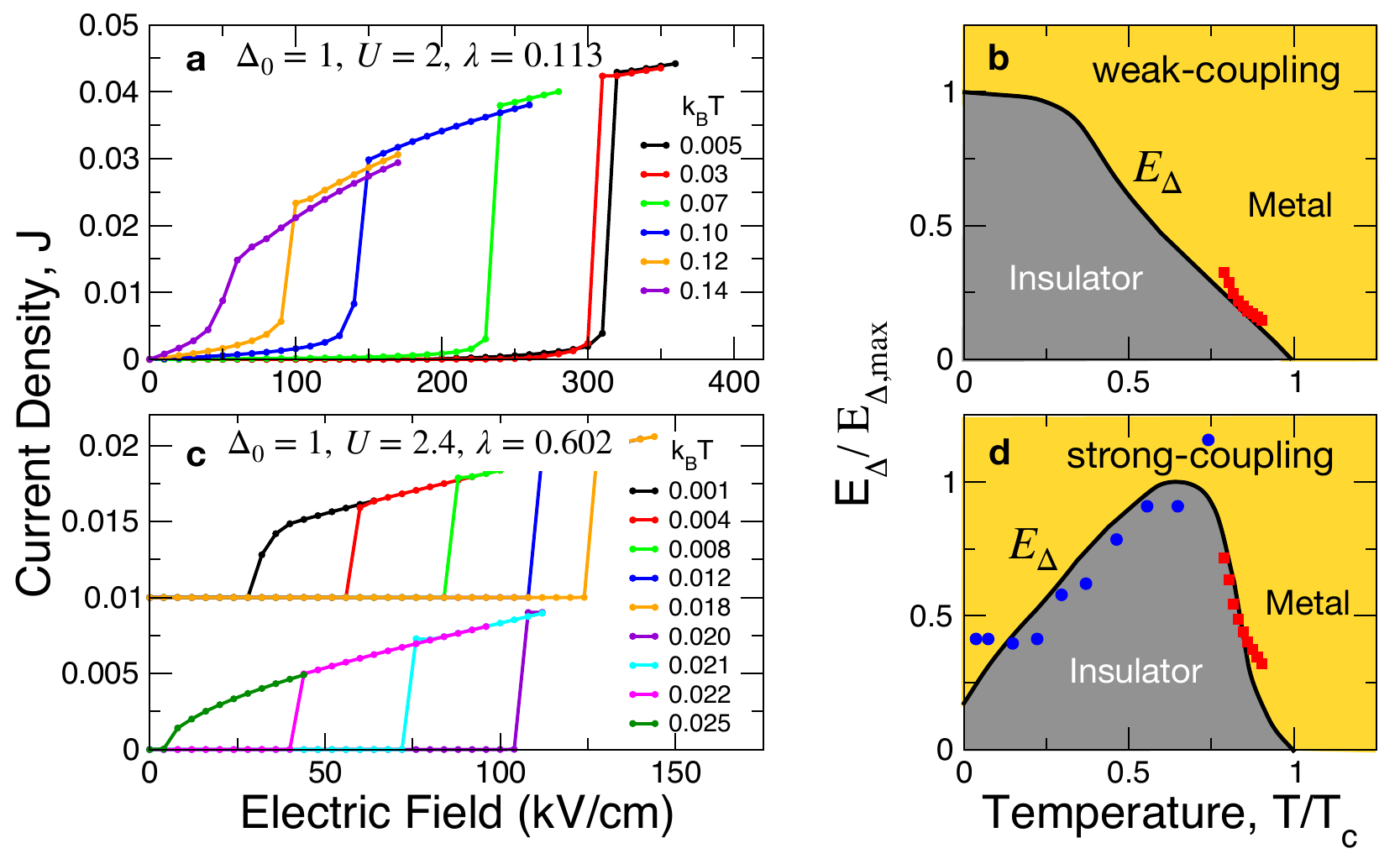}}}
\caption{{\bf Mott-like abrupt IMT a,} Current-voltage
($I-V$) relation in the limit of weak coupling to acoustic
phonons, displaying a single-stage discontinuous
insulator-to-metal transition (IMT). 
{\bf b,} Switching field versus temperature
($E_\Delta$ vs. $T$) phase diagram for (a) normalized to the maximum
of $E_\Delta$ and the transition temperature $T_c$, respectively. $k_BT_c=0.15$. The red squares
are switching fields of W$_x$V$_{1-x}$O$_2$ with $x=0.0114$ and
$T_c=280$ K (adapted from Ref~\cite{tlwu}, $y$-axis arbitrarily scaled).
{\bf c,} $I-V$ relation in the strong-coupling
limit showing non-monotonic dependence of $E_\Delta$ on $T$.
{\bf d,} Non-monotonic $E_\Delta$ versus $T$.
The initial increase of $E_\Delta$
at low $T$ demonstrates the non-thermal nature of the IMT in
the strong-coupling limit. $k_BT_c=0.025$. The blue
circles are switching fields of TiS$_3$ with
$T_c=290$ K (adapted from Ref~\cite{randleACSnano,randleAPL2022},
$y$-axis arbitrarily scaled). The red squres are the same as (b).
}
\label{fig4}
\end{center}
\end{figure}

Finally, let us discuss the case of electron coupling to acoustic
phonons, with a continuous spectrum of the form $\omega_k\propto k$ and
cutoff Debye energy chosen as $\omega_D= 0.6$ eV. Compared to optical
phonons, the influence of acoustic phonons on the nonequilibrium
dynamics is more dramatic.  We discuss this situation for both the weak
and strong limits of electron-phonon coupling, as shown in
Fig.~\ref{fig4}. In the weak-coupling limit ($\lambda=0.113$),
corresponding to panels (a) and (b), the IMT is dominated by a strongly
discontinuous collapse of the gap. While the signature of the avalanche
is still present as a precursor to the IMT, the range of the avalanche
region is much narrower than that found for coupling to optical phonons.
The avalanche current is also very small so that its effect is
insignificant. $E_\Delta$ decreases monotonically with
increasing temperature and the $E-T$ phase diagram is conventional
and fully consistent with the thermal switching
scenario. We tested this result against experimental data by overlaying
the experimental switching fields in the Vanadium oxides
W$_x$V$_{1-x}$O$_2$~\cite{tlwu} at temperatures close to the transition
temperature. In addition to an overall agreement of the trends, the
lightly concave curve shape~\cite{tlwu,maki1986} found in the experiment is 
reproduced by the theory.

In the strong electron-phonon coupling limit ($\lambda=0.602$), on the
other hand, $E_\Delta$ is strongly nonmonotonic,
increasing with $T$ in the low-$T$ limit. As is especially apparent from
panel (c), the intermediate field region between $E_T$ and
$E_\Delta$ has diminished dramatically so that the IMT
appears as a fully single-stage transition, as often observed in
correlated transition-metal
oxides~\cite{janod,stoliar,zimmers,guenon,htkim}. In this limit, the
strong low-energy excitations of acoustic phonons cause the IMT to bypass
the CDW-like pseudo-gap state. We therefore make a
prediction that the avalanche physics which controls the CDW threshold
field $E_T$ is manifested in Mott insulators in a switching field
$E_\Delta(T)$ that increases with $T$. We compare our results with the
single-stage switching fields measured in
TiS$_3$~\cite{randleACSnano,randleAPL2022} (blue circles). The agreement
between theory and experiment is quite reasonable. Altogether, the
$E_\Delta(T)$ behavior in the weak- and strong-coupling limits
demonstrates a crossover between the quantum and thermal switching
mechanisms as temperature is varied.

To conclude, we have established a quantum avalanche mechanism as a
generic scenario for the nonequilibrium phase transition in
correlated insulators. The proposed mechanism
not only resolves the long-standing discrepancy between the observed and
predicted switching fields in these materials, but also sheds new light
on the origins of the crossover between the quantum and thermal
scenarios in resistive transitions. The existence of the pseudo-gapped
metallic states may be directly verified experimentally by using
transient electrical pulses of varying duration to control the amount of
hot-phonon generation. While we have discussed this mechanism with
electron-phonon coupling, the avalanche may also arise via coupling to
other bosonic excitations, such as the Goldstone modes associated with
the order parameter responsible for the opening of the charge gap. Such
a scenario may provide a more direct path for the destruction of the
correlated gap. Here we have presented a minimal framework for the
quantum avalanche that is fundamentally different from the Landau-Zener
mechanism. The study of the avalanche with spatial inhomogeneity, and
its interplay with disorder, is left for future research.

\bigskip
\centerline{\large\bf Methods}

\bigskip
\textbf{Calculation of Self-Energy and Effective
Temperature: }
The many-body calculations in this work are based on the nonequilibrium Green's
function technique approximated by the DMFT
scheme~\cite{aoki,liprl2015,ligraphene,moire2023}, in which we limit the self-energy to be
diagonal in site and orbital indices. Full details on the
nonequilibrium DMFT are given in the Supplementary Information.
The electron and phonon self-energies, expressed respectively as
\begin{eqnarray}
\Sigma^\lessgtr_{{\rm ep},\alpha}({\bf r},\omega)
& = &
ig_{\rm ep}^2\int\frac{{\rm d}\omega'}{2\pi}{\cal G}^\lessgtr_{\alpha\alpha}({\bf
r},\omega-\omega')D^\lessgtr_\alpha(\omega'),\\
\Pi^\lessgtr_{{\rm ep},\alpha}(\omega)
& = &
-2ig_{\rm
ep}^2\int\frac{{\rm d}\omega'}{2\pi}G^\lessgtr_{\alpha\alpha}({\bf
r},\omega+\omega')G^\gtrless_{\alpha\alpha}({\bf r},\omega'),
\end{eqnarray}
are iterated to convergence. (The factor 2 in the phonon
self-energy accounts for spin degeneracy.) The
self-energies without the vertex correction, i.e., Migdal
approximation~\cite{schrieffer}, are reasonable since we are in the weak- to moderate
el-ph coupling limit, as can be seen with less than 10\% level shift of
the bandedge in Fig.~\ref{fig2}(b).

Once we achieve full convergence, we
compute the distribution functions as
\begin{equation}
f_{\rm el}(\omega)=-\frac12\frac{\sum_\alpha{\rm Im}G^<_{\alpha\alpha}({\bf
r}=0,\omega)}{\sum_\alpha{\rm
Im}G^R_{\alpha\alpha}({\bf r}=0,\omega)},\quad
n_{\rm ph}(\omega)=\frac12\frac{{\rm Im}D^<(\omega)}{{\rm
Im}D^R(\omega)}.
\end{equation}
We define the effective temperature for electrons and phonons in terms
of the first moment of the distribution, which correctly reduces to the bath temperature
in the equilibrium limit, as
\begin{eqnarray}
T_{\rm el}^2 & = &
\frac{6}{\pi^2}\int^\infty_{-\infty} \omega[f_{\rm
el}(\omega)-\Theta(-\omega)]{\rm d}\omega \nonumber
\\
T_{\rm ph}^2 & = &
\frac{6}{\pi^2}\int_0^\infty \omega n_{\rm ph}(\omega){\rm d}\omega,
\end{eqnarray}
with the step-function $\Theta(x)$. As shown in Fig.~\ref{fig2}{\bf b},
electronic distribution functions may deviate strongly from the
Fermi-Dirac form, in which cases $T_{\rm el}$ provides an
approximate measure of nonequilibrium energy excitation.

%
%

\bigskip
\centerline{\large\bf Acknowledgements}

\bigskip
JEH is grateful for computational support from the CCR at
Buffalo. JEH is supported by Air Force Office of Scientific Research under award no. FA9550-22-1-0349. CA acknowledges the support from the French ANR ``MoMA'' project ANR-19-CE30-0020 and from the project 6004-1 of the Indo-French Centre for the Promotion of Advanced Research (IFCPAR).
JHH acknowledges the support from the Institute for Basic Science in the Republic of Korea through the project IBS-R024-D1.
KSK was supported by the Ministry of Education, Science, and Technology 
(Grants No. NRF-2021R1A2C1006453 and No. NRF-2021R1A4A3029839) of the National Research Foundation of Korea (NRF) 
and by TJ Park Science Fellowship of the POSCO TJ Park Foundation.
We are much grateful to Sambandamurthy Ganapathy, Han-Woong Yeom, Emmanuel Baudin for  helpful discussions.

\end{document}